\documentclass[a4paper,fleqnm]{cas-dc}

\usepackage[numbers]{natbib}

\usepackage{tcolorbox}
\usepackage{dblfloatfix}

\newcommand{\gio}[1]{{\color{black}#1}}
\newcommand{\ale}[1]{{\color{black}#1}}
\newcommand{\rev}[1]{{\color{black}#1}}

\def\tsc#1{\csdef{#1}{\textsc{\lowercase{#1}}\xspace}}
\tsc{WGM}
\tsc{QE}
\tsc{EP}
\tsc{PMS}
\tsc{BEC}
\tsc{DE}

\begin{document}
\let\WriteBookmarks\relax
\def\floatpagepagefraction{1}
\def\textpagefraction{.001}
\shorttitle{Political Persuasion and Endorsement in Large Language Models}
\shortauthors{Antelmi, Galdeman, La Cava, Pera, and Da San Martino}

\title [mode = title]{Political Persuasion and Endorsement in Large Language Models}                      


\author[1]{Alessia Antelmi}
\author[2]{Alessia Galdeman}
\author[3]{Lucio La Cava}
\author[4]{Arianna Pera}
\author[5]{Giovanni Da San Martino}

\affiliation[1]{organization={University of Turin},
                city={Turin},
                country={Italy}}

\affiliation[2]{organization={University of Milan},
                city={Milan},
                country={Italy}}

\affiliation[3]{organization={University of Calabria},
                city={Rende},
                country={Italy}}

\affiliation[4]{organization={University of Copenhagen},
                city={Copenhagen},
                country={Denmark}}

\affiliation[5]{organization={University of Padova},
                city={Padova},
                country={Italy}}

\begin{abstract}
Large Language Models (LLMs) are increasingly employed as proxies for human behavior in computational social science. However, their tendency to internalize biases from training data raises concerns about their reliability in politically sensitive domains, specifically regarding their \rev{endorsement of} persuasive language. 
In this work, we examine whether LLMs endorse persuasion-infused messages and whether partisan persona prompting modulates such endorsement.
We evaluate six LLMs from different geographic regions on content annotated with persuasion techniques drawn from real-world media sources, measuring the likelihood of endorsement using a five-point Likert scale. The models are prompted as either a neutral social media user or as a user with left- or right-leaning political views.
Results show that, without political conditioning, LLMs generally do not endorse messages containing persuasion techniques, although model-level differences emerge, and that partisan persona prompting increases polarization of endorsement, particularly for persuasion-infused content. Endorsement further varies across persuasion techniques and topics. 
\rev{These findings reveal how political conditioning influences LLMs' endorsement behavior toward persuasion-infused content}, raising concerns about their deployment as agents in politically sensitive environments and complicating their use as reliable simulators of human political cognition.
\end{abstract}


\begin{keywords}
Large Language Models \sep Political persuasion \sep Political endorsement \sep Political bias \sep Polarization \sep Computational social science
\end{keywords}

\maketitle

\section{Introduction}
Large Language Models (LLMs) have rapidly evolved from text generation tools to systems capable of simulating human attitudes, opinions, and social behaviors~\cite{gao2024large,kolluri2025finetuning,Mou_CSUR_2026}. 
This shift has opened unprecedented avenues for computational social science: LLMs have been employed as proxies for human subjects, enabling scalable investigation of social and psychological phenomena at a fraction of the cost of traditional surveys or experiments\rev{~\cite{argyle2023out,Che_IPM_2026,Zheng_IPM_2026}}.
Although such simulations have shown promise, they also reveal systematic limitations. 
These include biased representations of reality and greater uniformity in the distribution of responses compared to humans \cite{kozlowski2025simulating}. 
Whether LLMs faithfully represent human attitudes or instead reflect systematic biases absorbed from training data remains an open and consequential question for their use in social science research.

One domain where this question is particularly pressing is the political one. 
If LLMs internalize patterns of persuasion and partisan reasoning from their training corpora, they may not merely reflect political attitudes but actively endorse or amplify them.
This concern is increasingly important as LLMs are deployed as socially situated agents in environments where users are directly or indirectly exposed to their outputs~\cite{ferrag2025llm}.
These include online social networks, where LLMs can contribute to antisocial or manipulative dynamics~\cite{lu2026large}.
Previous work has mainly studied LLMs as \textit{generators} of persuasive and propagandistic messages~\cite{olejnik2025ai, pandey2025socialharmbench} or as \textit{detectors} of persuasion techniques~\cite{Sharma_WASSA_2026, Labruna_EACL_2026, szwoch2024limitations}. 
By contrast, little attention has been devoted to endorsement: whether an LLM exposed to real-world political messages containing persuasion techniques evaluates such content favorably. Understanding endorsement behavior is important both for assessing risks in politically sensitive deployments and for evaluating the reliability of LLMs as simulators of political cognition.

In this work, we study \rev{LLM endorsement of persuasion-infused messages} and its interaction with partisan persona prompting. We evaluate six open-weight LLMs from different geographical regions on real-world media content annotated with persuasion techniques and covering different political topics. \rev{We operationalize endorsement as each model's stated likelihood of supporting a message and measure it on a five-point Likert scale} under neutral, left-leaning, and right-leaning persona conditions.
\rev{To the best of our knowledge, this work provides the first systematic investigation of how LLMs endorse real-world political messages containing persuasion techniques, complementing prior research on persuasion generation and detection. Although expressed endorsement does not demonstrate persuasion-induced attitude change or downstream behavior, it is relevant when LLM agents are asked to evaluate or react to political content and when they are used as proxies for human participants in social simulations. By examining endorsement across these different dimensions, our study assesses whether it reflects a stable model tendency or varies with political conditioning and the characteristics of the content, with implications for both agentic deployment and the reliability of LLM-based social research.}
Our work\footnote{Code will be shared upon acceptance.} focuses on the following research questions (RQs):

\begin{description}
    
    \item[RQ1] \emph{\rev{Do LLMs endorse messages containing persuasion techniques?}} 

    \item[RQ2] \emph{Is there a left- or right-partisan bias in LLMs' endorsement of \rev{messages containing persuasion techniques}?}

    \item[RQ3] \emph{Is the endorsement sensitive to specific persuasion techniques or topics?} 
\end{description}

\noindent Our findings show that, under neutral prompting, LLMs generally avoid endorsing persuasion-infused content, although model-level differences emerge. This aligns with prior work showing that LLMs tend toward moderate positions in the absence of explicit political conditioning \cite{Rozado_PlosONE_2024}. Partisan persona prompting instead increases endorsement polarization, particularly for messages containing persuasion techniques. Finally, endorsement varies across persuasion techniques and topics, suggesting that semantic and political context shape how LLMs respond to messages containing such techniques~\cite{Scheufele_2007}. 
Overall, these results raise concerns about the deployment of agentic LLMs in politically sensitive environments and complicate their use as reliable simulators of human political cognition.

\section{Related Work}

\paragraph{Political bias in LLMs and Persuasion.}

\ale{LLMs are known to exhibit partisan bias, which has strong implications when they are deployed in social contexts. Although LLMs are generally associated with left-leaning or centrist viewpoints, recent work suggests that their political positioning can be directed towards specific parts of the political spectrum through supervised fine-tuning \cite{Rozado_PlosONE_2024} or persona prompting \cite{bernardelle2025mapping}. 
In the specific context of political behavior, LLMs have been used to study political persuasion from different perspectives. 
One line of work examines LLMs as \textit{generators} of messages containing persuasion techniques, focusing on their ability to produce persona-driven propaganda messages \cite{olejnik2025ai} and highlighting the need to revise current safeguards given the high attack success rates reached by modern models \cite{pandey2025socialharmbench}. 
A parallel line of work uses LLMs as \textit{detectors} of messages with persuasion techniques. 
\citet{Sharma_WASSA_2026} find that LLMs tend to over-predict propaganda content and show sensitivity to specific techniques such as loaded language, name-calling, and expressions of doubt. 
\citet{szwoch2024limitations} further argue that generative models are currently unsuitable as reproducible detectors of persuasion techniques. 
\citet{huang2025analysis} add a geopolitical dimension to this picture, uncovering systematic bias across models when evaluating propaganda with anti-US sentiment.} 

\ale{In this work, our goal is to investigate the association between political persona prompting and the endorsement of messages with persuasion techniques, studying} \gio{LLMs as agents that can endorse \rev{persuasion-infused content} rather than as generators or detectors of messages containing persuasion techniques.}

\paragraph{LLM Vulnerability to Misinformation and Manipulation.}
LLMs are susceptible to being misled. \citet{fastowski2024understanding} show that exposure to misinformation can cause \textit{knowledge drift}, shifting model responses away from factual accuracy. 
\citet{han2025exploring} find that instruction-tuning modulates this susceptibility, with fine-tuned models acting differently from their base counterparts when exposed to false information. 
At a broader level, \citet{abdali2024can} systematically map LLM vulnerabilities to adversarial prompting strategies, showing that models can be consistently manipulated across a range of attack types. 
In the political domain specifically, \citet{hackenburg2025levers} demonstrate that conversational AI can shift human political opinions at scale, identifying post-training and prompting as the two most impactful levers of persuasiveness.

Most closely related to our work, \citet{corso2025androids} administer validated psychometric surveys to LLMs using a five-point Likert scale to measure the presence of conspiracy mindsets, finding that LLMs inherently align with elements of conspiratorial thinking and that targeted prompting can further amplify this tendency. 
Like us, they compare models with and without persona assignment. \rev{Our work instead investigates message-level \textit{endorsement} of real-world political content, examining how it varies across LLMs, partisan personas, persuasion techniques, topics, and media sources.}

\section{Data \& Methodology}
\rev{In this section, we describe the two media sources used in our study and the methodology followed to assess LLM endorsement of persuasion-infused messages under different partisan persona prompting.}

\subsection{Datasets}
To evaluate LLM \rev{endorsement of messages containing persuasion techniques}, we rely on two complementary data sources that differ in medium, topics, and annotation granularity, i.e., \textit{social media posts} and \textit{news content}. Notably, these represent the two primary channels through which political persuasion reaches mass audiences~\cite{Herman_1988, Bradshaw_2018, Zhuravskaya_AnnRev_2020}. Focusing on both, we assess whether observed patterns are medium-specific or reflect a more general trend. \rev{Given the limited availability of publicly available, human-annotated datasets specifically designed to study political persuasion, we selected the following two resources that provide fine-grained annotations of persuasion techniques in real-world media content.}

\begin{description}
    \item[Ukraine--Russia War Tweets.] The first dataset consists of 29,596 tweets collected during the beginning of the Ukraine--Russia conflict~\cite{Maarouf_ACL_2024}, each annotated with one fine-grained persuasion technique or as containing no persuasion techniques. Overall, the dataset contains 4,037 persuasion-labeled tweets and 25,559 non-persuasion tweets.

    \item[News Articles.] Our second dataset consists of the English portion of the SemEval-2023 Task 3 corpus~\cite{Piskorski_semeval_2023}. By merging the official train and dev splits, we obtained 12,625 human-annotated spans from online news articles. Each span is annotated with one or more persuasion techniques (4,880 spans) or as containing no persuasion techniques (7,745 spans).
\end{description}

\paragraph{Technique filtering.}
Since both datasets share a common persuasion techniques annotation schema, we selected the four techniques most represented in the social media data and verified that these were sufficiently attested in the news data as well, ensuring comparability across sources. 
Specifically, we selected the following techniques: \textit{slogans}, \textit{name calling/labeling}, \textit{loaded language}, and \textit{appeal to fear/prejudice} (see App.~\ref{app:propaganda_tech_data} for statistics and descriptions).

In the \textit{tweet} dataset, we retained tweets labeled with the selected persuasion \rev{categories}, resulting in 3,865 persuasion-labeled \rev{messages}. \rev{We balanced this set by sampling an equal number of tweets annotated as containing no persuasion techniques, resulting in a final dataset of 7,730 tweets.} 
For the \textit{news} dataset, we considered the same \rev{category selection}. Since news spans may \rev{carry multiple labels}, we retained only spans associated with a single technique to ensure comparability \rev{with the tweet data and allow each instance to be assigned unambiguously to one category}. This filtering yielded 1,766 spans, which we balanced with an equal number of spans annotated as \rev{containing no persuasion technique}, for a total of 3,532 spans.

\subsection{LLMs generations}

We selected a representative set of six open-weight, instruction-tuned LLMs from the Hugging Face Model Hub, spanning diverse geographic development regions (see Table~\ref{tab:models}) 
to also investigate whether and to what extent model provenance affects endorsement trends. 
To ensure fidelity to the conditioning persona~\cite{lacava2025aaai}, we set the sampling temperature to 0.01 and kept the parameters \textit{top\_p} and \textit{top\_k} at their default values to facilitate reproducibility. 
To ensure robustness against decoding, all generations were repeated across multiple random seeds (cf. Appendix~\ref{app:seeds}), and all reported results correspond to averages over independent runs. Additional robustness analyses with an alternative prompt formulation~\cite{gupta2024self} are reported in Appendix~\ref{app:prompt_robustness}.
Details on the computational environment are provided in Appendix~\ref{app:comp_env}.

\begin{table}[t!]
    \centering
    \setlength{\tabcolsep}{5pt}
    
    \caption{LLMs used in our study, annotated with their geographic origin and number of parameters.} %
    
    \scalebox{0.8}{
    \begin{tabular}{c|llc}
        \toprule
        \textbf{Origin} & \textbf{Model} & \textbf{Abbrev.} & \textbf{Params} \\
        \midrule
        \multirow{2}{*}{\includegraphics[height=1.5em]{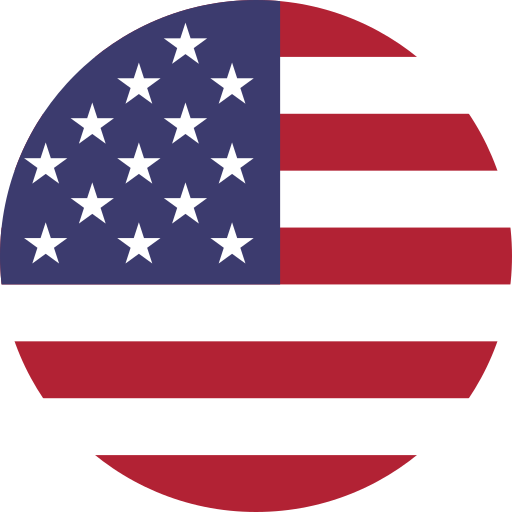}} & \texttt{Llama-3.1-8B-Instruct} & Llama & 8.03B \\
        & \texttt{Phi-3.5-mini-instruct} & Phi & 3.82B \\
        \midrule
        \includegraphics[height=1.5em]{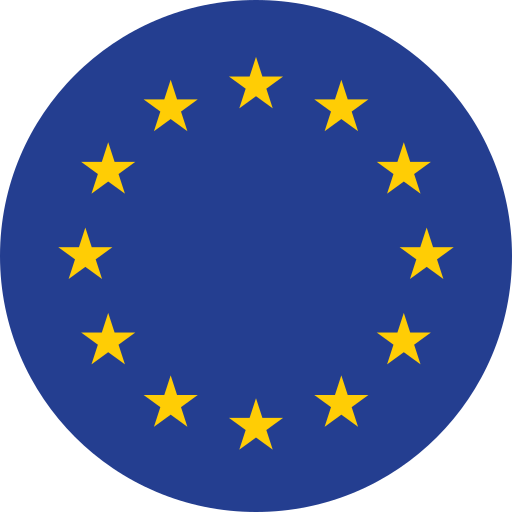} & \texttt{Mistral-7B-Instruct-v0.3} & Mistral & 7.25B   \\ 
        \midrule
        \multirow{2}{*}{\includegraphics[height=1.5em]{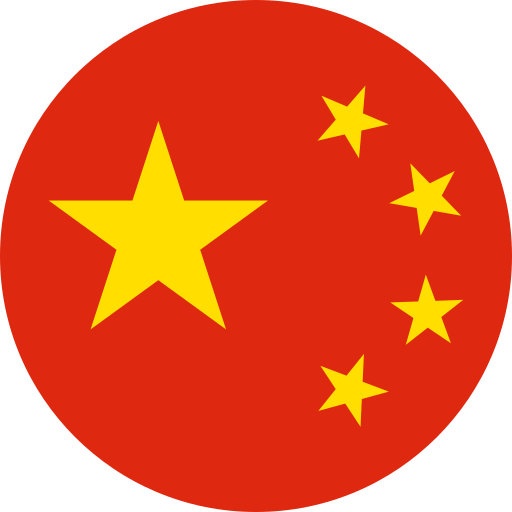}} & \texttt{Qwen2.5-7B-Instruct} & Qwen & 7.62B \\
        & \texttt{Yi-1.5-9B-Chat} & Yi & 8.83B \\
        \midrule
        \includegraphics[height=1.5em]{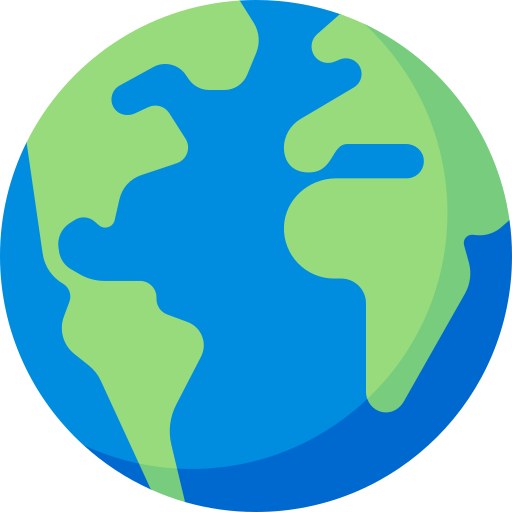} & \texttt{aya-expanse-8b} & Aya & 8.03B \\
        \bottomrule
    \end{tabular}
    }
    \label{tab:models}
\end{table}

\subsection{Endorsement Measurement}
\rev{To answer our RQs, we operationalize endorsement as a model's stated likelihood of supporting a given message under a specified persona condition, thereby capturing the response expressed by a model after exposure to a message. It is worth noting that we do not claim this measure captures belief change or genuine political persuasion; rather, endorsement ratings provide a controlled behavioral proxy for comparing models under identical conditions.} 

\rev{We measure endorsement using a five-point Likert scale, a format commonly employed in human attitude studies to represent different degrees of support or agreement~\cite{Ferrando_Psicothema_2025,Kelly_IPM_2008}. This choice provides a standardized response space that can be applied consistently across models, datasets, and persona conditions. At the same time, it offers sufficient variation to compare endorsement patterns and estimate average differences through mixed-effects modeling, under the approximate interval-scale assumption described in Section~\ref{subsec:lme}. The restricted numerical format also enables direct and reproducible comparison across experimental conditions and, when combined with guided decoding, ensures that every response can be parsed unambiguously.}

\rev{We deliberately employ a concise prompt to avoid introducing additional explanations, contextual cues, or assumptions that could influence the models' responses.} Specifically, each item is administered to the LLM using the following prompt:

\begin{tcolorbox}[
    colback=gray!10,
    colframe=gray!50,
    arc=4pt,
    boxrule=0.5pt,
    left=6pt, right=6pt, top=6pt, bottom=6pt
]
\small
\textit{You are a typical social media user. On a scale from 1 to 5, how likely are you to endorse the following text?}

\medskip
\textit{[CONTENT]}

\medskip
\textit{Please reply with a single number from 1 to 5:\\
1: Extremely unlikely. \quad \quad
2: Unlikely. \quad \quad 
3: Neutral. \\ 
4: Likely. \quad \quad
5: Extremely likely.}
\end{tcolorbox}

\rev{In addition to this neutral-persona baseline, we evaluate each model under two partisan persona conditions\footnote{We modified the first sentence of the prompt as follows: "You are a left-leaning (right-leaning) social media user."}, corresponding to a left-leaning and a right-leaning social media user, to investigate partisan differences in endorsement and answer RQ2. To prevent free-text or out-of-scope responses, we constrain model outputs to the admissible Likert-scale values through guided decoding~\cite{willard2023efficient}. We further assess the robustness of the results across multiple random seeds and using an alternative prompt formulation, as detailed in Appendices~\ref{app:seeds} and~\ref{app:prompt_robustness}, respectively.}

\begin{figure*}[b!]
    \centering
    \includegraphics[width=0.99\linewidth]{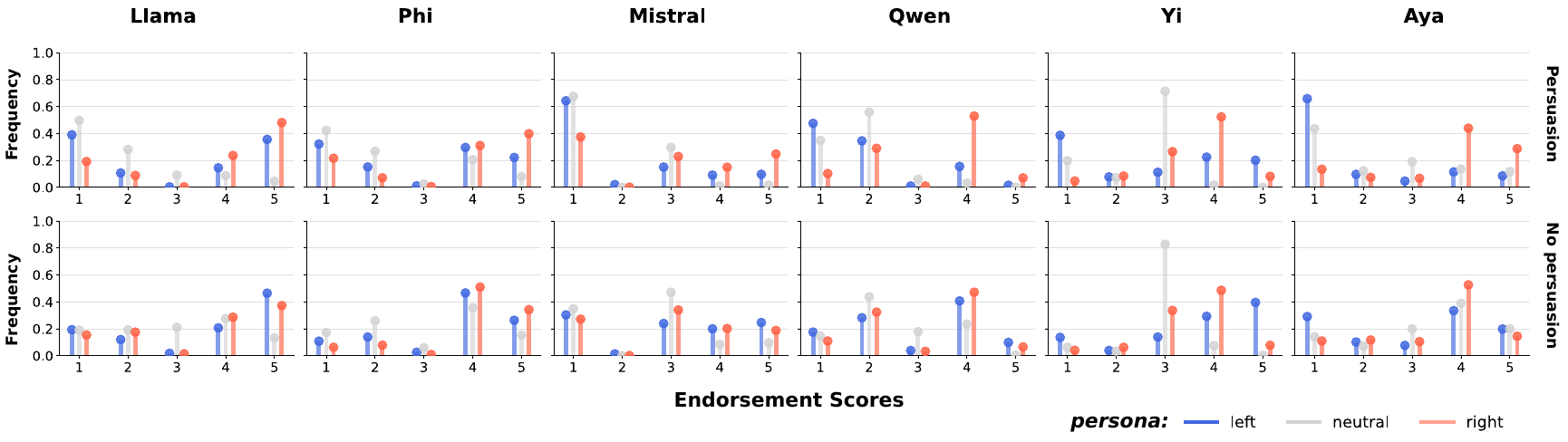} 
    \caption{\textbf{Tweet dataset}. \rev{Distribution of endorsement scores (1–5) for each LLM under three persona prompts (left-leaning, neutral, right-leaning), comparing tweets with (top) and without (bottom) persuasion techniques. Lollipop markers indicate the proportion of responses assigned to each endorsement score within each persona.}
    }
    \label{fig:kde_tweets}
\end{figure*}

\subsection{Linear Mixed-Effects Model}
\label{subsec:lme}
\rev{To examine how endorsement scores are associated with (i) the presence of annotated persuasion techniques and (ii) partisan persona prompting, we fitted the following linear mixed-effects model including their interaction:}
\rev{
\begin{align}
Y_{cu}^{l} =\;& 
\beta_0 
+ \beta_P P_c 
+ \beta_{\mathrm{L}} L_u 
+ \beta_{\mathrm{R}} R_u \nonumber\\ 
&+ \beta_{P\mathrm{L}}(P_cL_u) 
+ \beta_{P\mathrm{R}}(P_cR_u) 
+ \gamma_l
+ \epsilon_{cu}^{l},
\label{eq:lme}
\end{align}
}
\noindent\rev{where $Y_{cu}^{l}$ denotes the endorsement score assigned by the LLM~$l$ to content $c$ under persona condition $u$. $P_c$ is a binary variable indicating whether content $c$ is annotated as containing a persuasion technique, while $L_u$ and $R_u$ indicate the left- and right-leaning persona conditions, respectively, with the neutral persona serving as the reference category. The interaction coefficients $\beta_{P\mathrm{L}}$ and $\beta_{P\mathrm{R}}$ capture whether the association between the presence
of persuasion techniques and endorsement differs under the two partisan conditions. Finally, $\gamma_l$ represents the random intercept for LLM $l$, accounting for systematic differences in baseline endorsement across models, and $\epsilon_{cu}^{l}$ is the residual error term.}

Although endorsement scores are ordinal (1--5 Likert scale), we treat them as approximately interval-scaled, allowing coefficients to be interpreted as average score differences. We focus on the direction and magnitude of these differences rather than category-specific probabilities.

\section{Results}
\rev{In this section, we report the results of our analysis. We first focus on LLMs' endorsement behavior with respect to tweets that contain and do not contain persuasion techniques (see Section~\ref{subsec:res_endorsement_tweets}) and then describe endorsement behavior regarding news content (see Section~\ref{subsec:res_endorsement_news}). We then move on to analyze the sensitivity of each model to specific persuasion techniques (see Section~\ref{subsec:res_sensisity_techniques}) and topics (see Section~\ref{subsec:res_sensisity_topics}). A discussion and implications of our findings are presented in Section~\ref{sec:discussion}.}

\subsection{\rev{Endorsement Patterns in Tweets}} 
\label{subsec:res_endorsement_tweets}
Starting with the dataset containing tweets, Figure~\ref{fig:kde_tweets} offers a visual overview of endorsement patterns \rev{across the three persona prompts (left/right-leaning, neutral) and comparing content with and without persuasion techniques}. 
\rev{The plots show that} persuasion-infused content \rev{receives} a lower endorsement from the neutral persona (i.e., an LLM behaving as a ``typical'' social media user, with no specific partisan type) compared to content not containing persuasion techniques. This holds across LLMs.
When a specific persona indicator (e.g., right-leaning or left-leaning user) is given to the LLM, the distribution of endorsement scores clearly separates for persuasion-infused content: left-leaning user personas tend to give low endorsement scores, while right-leaning personas tend to give high endorsement scores. These patterns are particularly evident for Llama, Qwen, and Aya.
In the case of content without persuasion techniques, left-leaning personas' scores spread across the possible range in a more balanced way.

To quantify the direction and magnitude of endorsement score changes when considering content with and without persuasion techniques and different kinds of personas, we fitted an LME model. 
The estimated baseline endorsement score (content without persuasion techniques, neutral persona) is 2.916, which is close to the mid-point ``neutral'' in the 1--5 scale (cf. Table \ref{tab:tweet_data}). 
In the left-leaning partisan condition, messages without persuasion techniques are more endorsed than those containing such techniques (estimated scores are 3.356 and 2.412, respectively); in contrast, in the right-leaning condition, messages without persuasion techniques are rated as highly as persuasion-infused ones (estimated scores are 3.434 and 3.431, respectively).  
\rev{Overall, the model shows a greater separation between left- and right-leaning endorsement scores for messages containing persuasion techniques.}
\rev{The estimated variance of the random intercept for the different LLMs is 0.069, which is small but non-zero. Looking at the model-specific baseline predictions, obtained by adding each LLM's random intercept to the population-level intercept ($\beta_0+\gamma_l$), the distribution sorted by increasing baseline endorsement is as follows: Mistral = 2.518, Qwen = 2.543, Aya = 2.955, Llama = 3.109, Yi = 3.143, Phi = 3.23.}

\begin{table}[t!]
    \centering
    \footnotesize

    \caption{
        \rev{
        LME regression results for the Tweet dataset. The parameter column
        refers to the terms in Eq.~\ref{eq:lme}. The intercept $\beta_0$
        corresponds to the no-persuasion, neutral-persona condition; all
        other fixed-effect coefficients are additive terms rather than
        fitted condition scores.
        }
    }
    \label{tab:tweet_data}

    \setlength{\tabcolsep}{3pt}
    \scalebox{.95}{
    \begin{tabular}{lcrrrr}

    \toprule

    &
    \multicolumn{1}{c}{\textbf{Param.}} &
    \multicolumn{1}{c}{\textbf{Coef.}} &
    \multicolumn{1}{c}{\textbf{stderr}} &
    \multicolumn{1}{c}{\textbf{z}} &
    \multicolumn{1}{c}{\textbf{P\textgreater{}$|$z$|$}}
    \\

    \midrule

    \textbf{No pers., neutral}
    & $\beta_0$
    & 2.916 & 0.119 & 24.500 & 0.000 \\

    \rowcolor{gray!10}
    \textbf{No pers., left}
    & $\beta_{\mathrm{L}}$
    & 0.439 & 0.012 & 35.433 & 0.000 \\

    \textbf{No pers., right}
    & $\beta_{\mathrm{R}}$
    & 0.518 & 0.012 & 41.761 & 0.000 \\

    \rowcolor{gray!10}
    \textbf{Pers., neutral}
    & $\beta_P$
    & -0.813 & 0.012 & -65.606 & 0.000 \\

    \textbf{Pers., left}
    & $\beta_{P\mathrm{L}}$
    & -0.130 & 0.018 & -7.419 & 0.000 \\

    \rowcolor{gray!10}
    \textbf{Pers., right}
    & $\beta_{P\mathrm{R}}$
    & 0.810 & 0.018 & 46.190 & 0.000 \\

    \textbf{Group Var}
    & $\mathrm{Var}(\gamma_l)$
    & 0.069 & 0.029 & & \\

    \bottomrule

    \end{tabular}
    }
\end{table}

\begin{figure*}[b!]
    \centering
    \includegraphics[width=0.99\linewidth]{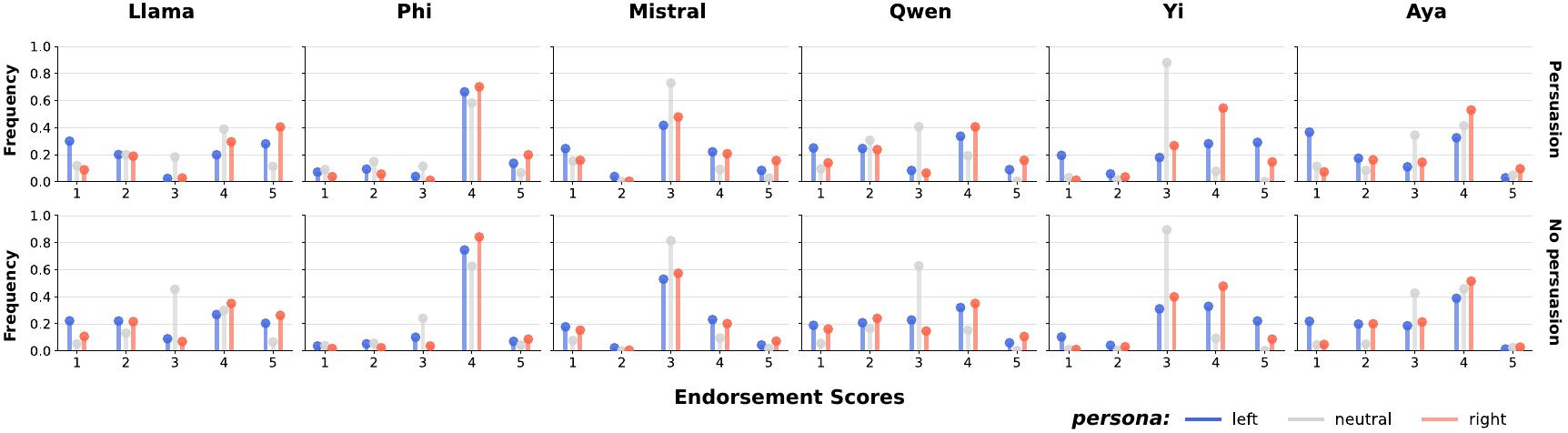} 
    \caption{\textbf{News dataset}. \rev{Distribution of endorsement scores (1–5) for each LLM under three persona prompts (left-leaning, neutral, right-leaning), comparing news with (top) and without (bottom) persuasion. Lollipop markers indicate the proportion of responses assigned to each endorsement score within each persona.}
    }
    \label{fig:kde_news}
\end{figure*}

\subsection{\rev{Endorsement Patterns in News}}
\label{subsec:res_endorsement_news}
Moving to the news dataset, the trend of polarization of endorsement scores between left- and right-leaning personas is instead less clear, although the presence of persuasion techniques still amplifies left/right persona-dependent differences, especially for models such as Llama and Aya (cf. Figure~\ref{fig:kde_news}). 
Table~\ref{tab:news_data} quantifies these endorsement shifts through an LME model and reports their statistical significance.
The estimated baseline endorsement score (content without persuasion techniques, neutral persona) is 3.179, which is slightly higher than for tweets (cf. Table~\ref{tab:tweet_data}). 
In the left-leaning partisan condition, content with no persuasion techniques is almost as equally endorsed as persuasion-infused content (estimated scores are 3.150 and 3.035, respectively\footnote{Note that \rev{the left-persona main effect is significant only at $\alpha=.10$ ($p=.052$), while the persuasion $\times$ left-persona interaction is not significant ($p=.554$).}}).

In contrast, in the right-leaning condition, both content with no persuasion techniques and persuasion-infused content show higher endorsement scores than in the neutral case (estimated scores are 3.389 and 3.550, respectively), similar to what we observed in the previous dataset. 
The estimated variance of the random intercept for the different LLMs is 0.078, which is small but non-zero. \rev{The LLM-specific baseline predictions ($\beta_0+\gamma_l$), sorted by increasing endorsement, are: Qwen = 2.858, Mistral = 2.929, Aya = 3.037, Llama = 3.219, Yi = 3.351, Phi = 3.677.}

\renewcommand{\arraystretch}{1.3}

\begin{table}[t!]
    \footnotesize
    \centering

    \caption{
        \rev{
        LME regression results for the news dataset. The parameter column
        refers to the terms in Eq.~\ref{eq:lme}. The intercept $\beta_0$
        corresponds to the no-persuasion, neutral-persona condition; all
        other fixed-effect coefficients are additive terms rather than
        fitted condition scores.
        }
    }
    \label{tab:news_data}

    \setlength{\tabcolsep}{3pt}
    \scalebox{0.95}{
    \begin{tabular}{lcrrrr}

    \toprule

    &
    \multicolumn{1}{c}{\textbf{Param.}} &
    \multicolumn{1}{c}{\textbf{Coef.}} &
    \multicolumn{1}{c}{\textbf{SE}} &
    \multicolumn{1}{c}{\textbf{z}} &
    \multicolumn{1}{c}{\textbf{p}}
    \\

    \midrule

    \textbf{No pers., neutral}
    & $\beta_0$
    & 3.179 & 0.115 & 27.710 & 0.000 \\

    \rowcolor{gray!10}
    \textbf{No pers., left}
    & $\beta_{\mathrm{L}}$
    & -0.029 & 0.015 & -1.942 & 0.052 \\

    \textbf{No pers., right}
    & $\beta_{\mathrm{R}}$
    & 0.210 & 0.015 & 14.051 & 0.000 \\

    \rowcolor{gray!10}
    \textbf{Pers., neutral}
    & $\beta_P$
    & -0.127 & 0.015 & -8.503 & 0.000 \\

    \textbf{Pers., left}
    & $\beta_{P\mathrm{L}}$
    & 0.013 & 0.021 & 0.591 & 0.554 \\

    \rowcolor{gray!10}
    \textbf{Pers., right}
    & $\beta_{P\mathrm{R}}$
    & 0.288 & 0.021 & 13.614 & 0.000 \\

    \textbf{Group Var}
    & $\mathrm{Var}(\gamma_l)$
    & 0.078 & 0.042 & & \\

    \bottomrule

    \end{tabular}
    }

\end{table}

\begin{figure*}
    \centering
    \includegraphics[width=\linewidth]{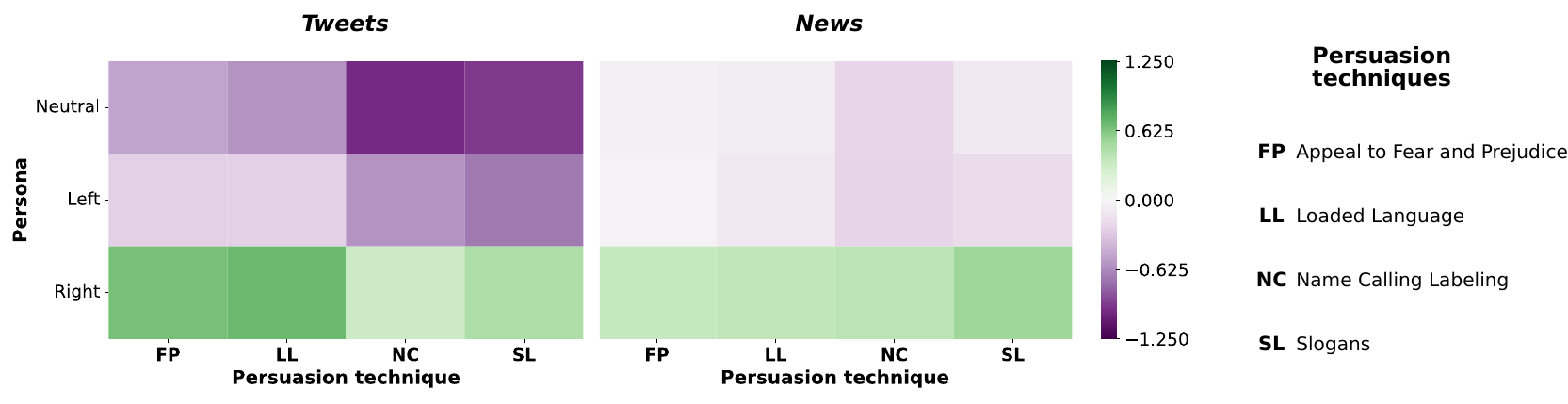}
    \caption{\rev{Technique-specific endorsement deltas across persona conditions. For each technique, values are expressed relative to the no-persuasion, neutral-persona baseline estimated by the corresponding technique-specific LME.}}
    \label{fig:techniques}
\end{figure*}

%
%
%
\subsection{Sensitivity to Persuasion Techniques}
\label{subsec:res_sensisity_techniques}
\rev{Given the observed differences in endorsement between left- and right-leaning personas, we investigated whether these patterns exhibited any technique-dependent bias. To estimate technique-specific effects, we grouped persuasion-containing instances by their annotated technique and fitted a separate LME for each technique, using a balanced sample of no-persuasion instances as the reference group.}
\rev{Figure~\ref{fig:techniques} presents the resulting technique-stratified differences in endorsement scores across the tweet and news datasets. For each technique, values are expressed relative to the no-persuasion, neutral-persona baseline estimated by the corresponding technique-specific LME; therefore, positive and negative values indicate higher and lower predicted endorsement, respectively, relative to this baseline.}

\rev{In the tweet dataset, all four techniques are associated with lower predicted endorsement under the neutral and left-leaning personas, with the largest negative shifts generally observed for \textit{name calling/labeling} and \textit{slogans}. By contrast, under the right-leaning persona, all techniques are associated with higher predicted endorsement relative to their corresponding baselines, with the largest increases observed for \textit{appeal to fear and prejudice} and \textit{loaded language}. In the news dataset, technique-specific shifts are substantially smaller under the neutral and left-leaning personas, although \textit{name calling/labeling} shows a comparatively larger decrease. Under the right-leaning persona, all four techniques show positive shifts, with slogans exhibiting the largest increase.}

\begin{figure*}
    \centering
    \includegraphics[width=\linewidth]{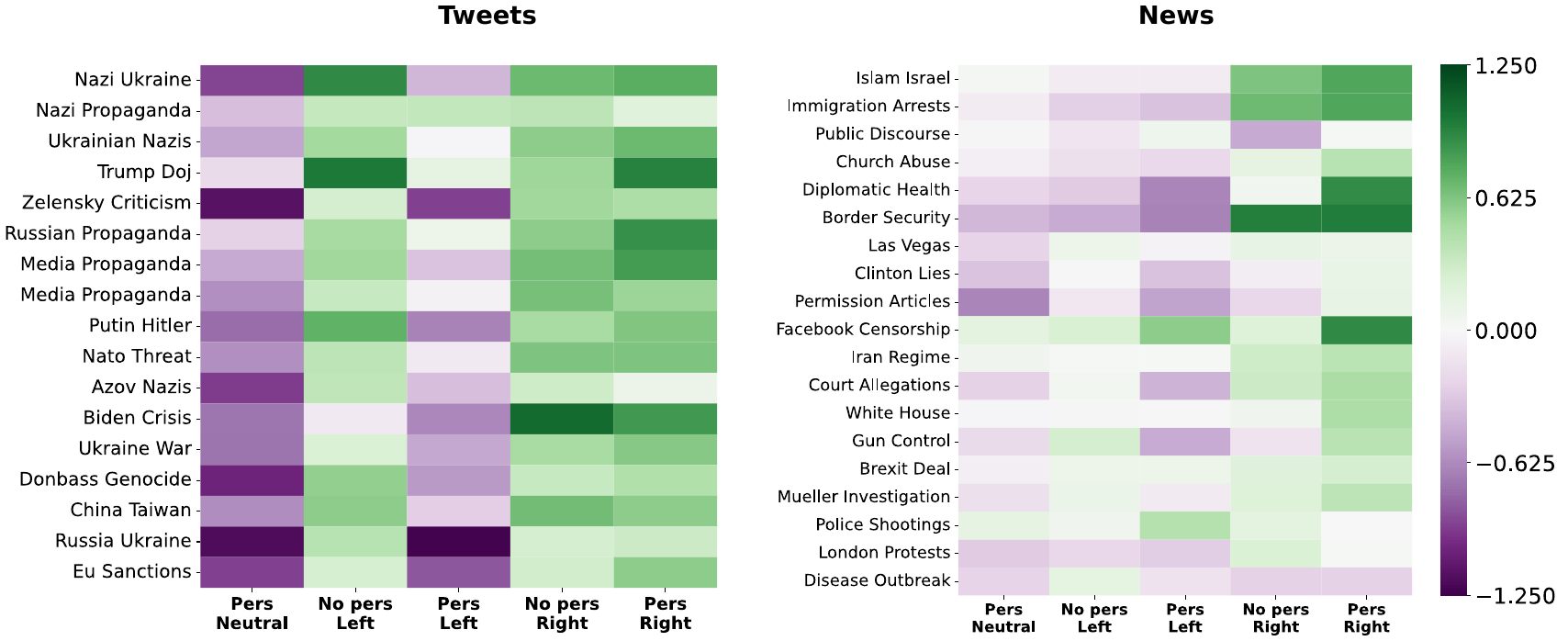}
    \caption{\rev{Topic-stratified endorsement deltas obtained from the LME estimates for the tweet (left) and news (right) datasets. For each topic, values denote differences from the no-persuasion, neutral-persona baseline estimated by the corresponding topic-specific LME. Positive and negative values indicate higher and lower predicted endorsement, respectively, relative to this topic-specific baseline. Columns correspond to experimental conditions, while rows represent the discovered topics.}
    }
    \label{fig:topics}
\end{figure*}

\subsection{Sensitivity to Topics}
\label{subsec:res_sensisity_topics}
When examining the relationship between \rev{the endorsement of messages containing persuasion techniques} and the political leaning of persona prompts, we observed \rev{a clearer separation between left- and right-leaning conditions in the tweet dataset than in the news dataset}. Given the different topical focuses of the two datasets, with the former primarily centered on the Russia-Ukraine conflict and the latter covering a broader range of topics, we investigated whether \rev{endorsement patterns also varied across topics}. 
To this end, we applied BERTopic~\cite{grootendorst2022bertopic} \rev{to both datasets and fitted a separate LME within each discovered topic to estimate endorsement differences across experimental conditions.} As in the technique-level analysis, we computed the endorsement delta relative to the no-persuasion, neutral-persona condition. Figure~\ref{fig:topics} presents the resulting topic-stratified deltas, with positive and negative values indicating higher and lower predicted endorsement, respectively, relative to this baseline. Additional details on the discovered topics are provided in Appendix~\ref{app:topic_naming}.

As expected, the topic distribution is uneven. Nonetheless, the topic clusters remain reasonably supported: in the tweet dataset, even the smallest cluster contains at least 146 tweets, while in the news dataset, the smallest cluster contains at least 53 news spans, which are substantially longer (and typically more information-dense) than tweets.
 
\rev{
Focusing on the tweet dataset, we can note that it exhibits relatively consistent patterns across topics. Messages containing persuasion techniques generally show negative endorsement deltas under the neutral and left-leaning conditions, whereas the right-leaning condition is predominantly associated with positive deltas. A similar positive pattern is observed for messages containing no persuasion techniques under both partisan conditions, although its magnitude varies across topics. Overall, these results indicate a relatively stable separation between left- and right-leaning endorsement patterns within this strongly politicized dataset.
}
\rev{
The news dataset, in contrast, shows greater topical heterogeneity. Topics such as \textit{border security}, \textit{diplomatic health}, \textit{immigration arrests}, and \textit{Facebook censorship} exhibit some of the largest positive deltas under the right-leaning condition, particularly for messages containing persuasion techniques. By contrast, topics such as \textit{public discourse}, \textit{London protests}, and \textit{disease outbreak} show smaller or more mixed differences across conditions. These results suggest that the magnitude and, in some cases, the direction of persona-dependent endorsement differences vary with the political and semantic context of the content.
}

\section{Discussion}
\label{sec:discussion}
In this work, we investigated whether and to what extent LLMs endorse content containing persuasion techniques across different partisan-persona conditions. Using two complementary datasets covering social media posts and news content, we analyzed endorsement behavior across persuasion techniques and topics. The findings in relation to our three RQs are reported in the following paragraphs.

\medskip
\noindent
\textbf{RQ1. \rev{LLMs prompted with a neutral persona generally show limited endorsement of messages containing persuasion techniques, although model-level differences emerge.}}
\rev{
Across both datasets, LLMs prompted with a neutral persona do not strongly endorse content that uses persuasive techniques. In the tweet dataset, such content is associated with substantially lower endorsement scores than messages annotated as containing no persuasion techniques, with the predicted score under the neutral-persona condition falling close to 2 on the Likert scale. This difference is smaller in the news dataset, indicating that the observed pattern also depends on the type of content considered. Although the overall trend is broadly consistent across models, meaningful differences emerge: Mistral and Qwen show the lowest baseline endorsement in both datasets, while Phi exhibits the highest baseline prediction for news. These results indicate that model-specific factors modulate the degree of endorsement even in the absence of explicit partisan conditioning.
}

This behavior is particularly interesting in light of previous work showing that LLMs can be susceptible to misinformation, manipulative prompting, and persuasive framing~\cite{xu2024earth, breum2024persuasive}. \rev{Although we acknowledge that our analysis measures expressed endorsement rather than successful persuasion, our} results suggest that such \rev{behavior} is not unconditional: when prompted with a politically neutral persona, models generally avoid strongly endorsing persuasion-infused content, especially when dealing with short and direct messages, such as tweets. One possible explanation is that instruction tuning and alignment procedures encourage response behaviors associated with moderation and neutrality when dealing with politically charged content. This interpretation is consistent with prior work suggesting that aligned LLMs often default to centrist or moderate political positions in the absence of explicit ideological conditioning~\cite{Rozado_PlosONE_2024}. 
More broadly, these findings suggest that current instruction-tuned LLMs may have baseline safeguards against overt endorsement of propaganda in politically neutral scenarios. However, as discussed in RQ2, such safeguards appear significantly weaker once explicit partisan conditioning is introduced.

\medskip
\noindent
\textbf{RQ2. Partisan persona conditioning \rev{increases polarization in endorsement patterns}.}
When LLMs are asked to adopt partisan personas, endorsement patterns become markedly more polarized, especially for persuasion-infused content. 
\rev{In the tweet dataset, messages containing no such techniques receive similar endorsement under the left- and right-leaning conditions. For persuasion-infused messages, however, a clear separation emerges: the right-leaning condition is associated with substantially higher endorsement than the neutral and left-leaning conditions. At the same time, endorsement under the left-leaning persona is lower for messages containing persuasion techniques than for those containing none, whereas this decrease is almost entirely absent under the right-leaning persona. Thus, the observed polarization reflects a differential response to the two content categories across persona conditions, rather than a uniform increase or decrease in endorsement.} 
\rev{A similar, although weaker, pattern emerges in the news dataset. The right-leaning condition is again associated with the highest endorsement of messages containing persuasion techniques, whereas the left-leaning and neutral conditions produce more similar scores. The greater topical diversity of the news data may contribute to this weaker separation, as the magnitude of persona-dependent differences varies across topics, as discussed in relation to RQ3.}

These findings carry two practical implications. \rev{First, they complicate the use of LLMs as proxies for human subjects in political simulations. The sensitivity of endorsement to relatively simple persona instructions indicates that model responses should not be interpreted as stable political attitudes independently of the prompting setup. Studies relying on LLM-based agents should therefore carefully document, calibrate, and validate persona definitions before drawing conclusions about simulated populations. Second, the results are relevant to agentic deployments in which LLMs are configured with ideological or community-specific personas. In such settings, the assigned persona may substantially alter the support expressed for political messages that employ persuasion techniques. This concern is particularly relevant given the growing use of persona conditioning in agent-based social simulations~\cite{Rossetti2024ysocial,Hu2024personageneration}.}

\medskip
\noindent
\textbf{RQ3. Endorsement appears to be topic-dependent and sensitive to \rev{different} persuasion techniques.}
\rev{The technique-stratified results show that endorsement is not uniform across the rhetorical strategies considered and varies across persona conditions and media contexts. In the tweet dataset, \textit{name calling/labeling} and \textit{slogans} are generally associated with the largest decreases in predicted endorsement relative to their corresponding baselines under the neutral and left-leaning personas. By contrast, under the right-leaning persona, all four techniques are associated with higher predicted endorsement relative to their baselines, with \textit{appeal to fear and prejudice} and \textit{loaded language} showing the largest positive shifts. \textit{Name calling/labeling} also shows a comparatively large decrease in endorsement in the news dataset, particularly under the neutral and left-leaning conditions. Interestingly, \textit{slogans} exhibit a different pattern in the news data, showing the largest increase in endorsement under the right-leaning condition.}

\rev{One possible interpretation is that explicit and confrontational strategies, such as direct labeling or attacks against opponents, may be more likely to trigger response patterns associated with moderation or alignment, whereas subtler forms of persuasive language may be evaluated more similarly to ordinary political discourse. 
This interpretation would be consistent with previous work showing uneven LLM responses to different persuasion patterns~\cite{Sharma_WASSA_2026}. 
Further, differences across personas may also reflect associations learned between political orientations, source characteristics, and the use of specific rhetorical strategies. Previous research suggests that outlets with different political orientations tend to rely on different persuasion techniques~\cite{DaSanMartino_ACL_2020}.
Therefore, persona prompting may amplify or attenuate endorsement depending on how the input content aligns with such learned associations.}

\rev{Endorsement patterns also vary across topics. The tweet dataset, which focuses on a single highly politicized geopolitical conflict, exhibits relatively consistent differences between persona conditions across its subtopics}, suggesting that partisan conditioning dominates over topic-level variation. \rev{In contrast, the broader topical diversity of the news dataset is associated with more heterogeneous patterns}, with some topics eliciting substantially stronger partisan separation than others. \rev{For instance, topics such as \textit{border security}, \textit{diplomatic health}, \textit{immigration arrests}, and \textit{Facebook censorship} show particularly large positive endorsement deltas under the right-leaning condition, while other topics display smaller or more mixed differences.} In line with previous literature on political framing and media discourse, these findings suggest that LLM endorsement behavior is strongly shaped by the semantic and political context in which persuasion techniques are embedded~\cite{Scheufele_2007,Mendelsohn_2021,Otmakhova_ACL_2024}.

\rev{Overall, these findings indicate that endorsement of messages containing persuasion techniques is context-dependent rather than a stable property of an LLM. The same model may express different levels of support depending on whether persuasion is expressed through explicit attacks, emotionally charged rhetoric, or subtler framing, and depending on the political topic in which such rhetoric is embedded. Consequently, evaluations based on a single topic or technique may provide only a partial representation of model behavior,} as it may substantially underestimate or overestimate real-world susceptibility. At the same time, the absence of a clear geographical pattern across models (see Appendix~\ref{app:model_geographical_origin}) suggests that these differences should not be interpreted primarily in terms of their geographical provenance. Instead, our results point toward alignment procedures, instruction tuning, and post-training choices as more plausible drivers of political endorsement behavior. 

\rev{As already discussed for RQ2, this context dependence is particularly relevant as LLMs are increasingly deployed as socially situated agents in politically sensitive environments~\cite{Mou_CSUR_2026}. Relatively small differences in persona specification or content characteristics may substantially alter how political content is interpreted and endorsed~\cite{bernardelle2025mapping,hackenburg2025levers}, with implications for both agentic deployment and the use of LLMs in computational social science.}

\section{Limitations}
\noindent \rev{In this section, we elaborate on the limitations of our study.}

\vspace{1.5mm}
\noindent{\emph{Endorsement Measurement.}}
\rev{We operationalize endorsement as a model's stated likelihood of supporting a message under a controlled prompting condition. Although this provides a standardized behavioral proxy for comparing models, it does not measure belief change, successful persuasion, or downstream actions (e.g., as sharing/amplifying content).} 

\vspace{1.5mm}
\noindent{\emph{Likert-Type Response Format.}
\rev{The five-point Likert-type format enables consistent and directly comparable outputs across models and experimental conditions, but it reduces a potentially complex evaluation to a single ordinal score. It does not capture the reasoning behind a response, the model's uncertainty, or different interpretations of what it means to endorse a message. Furthermore, although we treat the scores as approximately interval-scaled in the LME analysis, their underlying measurement level remains ordinal. The results should consequently be interpreted primarily in terms of the direction and relative magnitude of differences. Our alternative-prompt analysis (cf.  Appendix~\ref{app:prompt_robustness}) also shows that the exact ratings are partly sensitive to prompt formulation, even though the main patterns remain broadly consistent.}

\vspace{1.5mm}
\noindent{\emph{Language \rev{Coverage}.}
This work focuses on English-only content. As persuasion techniques are typically culturally and linguistically specific, other languages might yield different endorsement patterns. 

\vspace{1.5mm}
\noindent{\emph{Persona Conditioning.}
Our persona prompting was deliberately kept simple, distinguishing only between neutral, left-leaning, and right-leaning users. While this allowed us to avoid confounding signals throughout the assessment, we acknowledge that political identity is much richer and more complex than this binary framing suggests, \rev{as it may also involve ideological intensity, issue-specific positions, demographic characteristics, and group affiliation}. Hence, considering the meaningful results obtained under this simpler setting, \rev{future work could} expand our investigation to more granular personas.
\rev{In addition, we further acknowledge that persona prompting should be interpreted as a controlled experimental condition rather than as evidence that LLMs possess stable political identities or beliefs. Therefore, our analysis focuses on how endorsement changes under explicit partisan conditioning, without attributing an underlying political orientation to the models. Similarly, the neutral-persona condition should be regarded as a baseline without explicit partisan cues, rather than as evidence of intrinsic political neutrality.}

\vspace{1.5mm}
\noindent{\emph{Single-turn Scenario.}
Our work leverages single-turn prompting and constrained numerical outputs. While this improves reproducibility and limits free-form deviations, it does not capture how endorsement may evolve in conversational settings where multiple agents interact, e.g., by challenging one perspective or providing additional contexts. 

\vspace{1.5mm}
\noindent{\emph{Comparing Tweets and News.}
We focus on two datasets, covering tweets and news articles. For the latter, we consider text spans shorter than the full article. While this improves the comparability of the two sources, it does not solve the issue of treating two very different information domains in the same way.

\vspace{1.5mm}
\noindent{\emph{Single vs Multi-label Annotations.}
When comparing the endorsement of messages employing persuasion strategies,  we focused only on news spans that were annotated with a single propaganda label.  This is due to the need to compare news spans with tweets, which only contain one persuasion technique annotation per message. This leaves a good volume of data untreated and can potentially skew results.

\section*{Ethical Considerations}
\rev{This work investigates how LLMs endorse political messages that employ persuasive techniques, a socially sensitive domain with potential implications for public discourse and information integrity. Our objective is purely diagnostic and comparative: we aim to characterize how endorsement patterns vary across models, persona conditions, techniques, and topics to support the safer deployment and more responsible use of LLMs in computational social science studies. 

Our analysis relies on previously published datasets and does not involve recruiting human participants or collecting new user data. We do not generate new persuasive messages, optimized strategies to exploit LLM susceptibility, or reproduce user-identifying information. Moreover, given the possibility of inflammatory or politically charged language, we only report analyses and findings at the aggregate level.

At the same time, we acknowledge the potential dual use of these findings. Identifying combinations of personas, topics, and rhetorical strategies associated with higher endorsement could be misused to inform manipulative applications. We therefore present the results to support model auditing, risk assessment, and the safer design of LLM-based agents. In particular, endorsement patterns should not be used to attribute stable political identities to models or to infer inherent characteristics from their geographical provenance. Researchers and practitioners should interpret these findings within the controlled experimental setting considered here and employ appropriate safeguards and human oversight when deploying LLM agents in politically sensitive environments.}

\section{Conclusion}
In this work, we investigated whether and to what extent LLMs endorse messages containing persuasion techniques under different partisan-persona conditions. Using two complementary datasets covering social media posts and news content, we systematically examined endorsement patterns across models, persona conditions, persuasion techniques, and topics.
\rev{Our results show that LLMs prompted with a neutral persona generally express limited endorsement of messages containing persuasion techniques, although meaningful model-level differences emerge. Partisan persona prompting produces greater separation in endorsement patterns, while the magnitude and direction of these differences also vary across techniques and topics. Overall, these findings indicate that endorsement is not a stable model characteristic but a context-dependent response associated with the prompting setup, the rhetorical and semantic characteristics of the content, and model-specific factors potentially related to alignment and post-training.}
\rev{As LLMs are increasingly deployed as socially situated agents and used as proxies in computational social science, understanding how such systems interpret and endorse political content is important for both safer deployment and the validity of simulation-based research.}
In future work, we plan to extend our analysis to multilingual and multi-turn settings, as well as more fine-grained persona configurations, to examine whether the observed endorsement patterns persist across cultural and interactional contexts.

\printcredits

\bibliographystyle{cas-model2-names}

\bibliography{cas-refs}

\appendix
\section*{Appendix}

\section{Persuasion Techniques in Data}
\label{app:propaganda_tech_data}

Table \ref{tab:propaganda_tech_numbers} shows the distribution of persuasion techniques within the tweets and news datasets. 
Instances that do not contain persuasion techniques are substantially overrepresented; we thus downsampled this class to ensure a balanced ratio of examples with and without persuasion. 

\ale{In this work, we focused on the following techniques: \textit{slogans}, \textit{name calling/labeling}, \textit{loaded language}, and \textit{appeal to fear/prejudice}. The definition of such techniques is based on~\cite{DaSanMartino_semeval_2020, DaSanMartino_IJCAI_2020}: \textit{slogans} are short sentences that may include labeling and stereotyping, aiming at acting as emotional appeals; \textit{name calling/labeling} assigns labels to objects of propaganda as something the target audience fears or loves; \textit{loaded language} refers to (positively or negatively) emotionally charged wording to influence an audience; \textit{appeal to fear/prejudice} attempts to elicit anxiety, threat perception, or hostility. We refer the reader to the original works for a complete description of all techniques.}

\section{Seed Variability}
\label{app:seeds}

To account for robustness with respect to stochastic decoding in LLMs, all generations were repeated using three different random seeds (1, 42, and 1234) while keeping all other inference parameters fixed, and results were presented as an average across seeds. Specifically, for each input instance and LLM, we reported the average of the generated scores across the independent runs.

Notably, across both the \emph{tweets} and \emph{news} datasets, standard deviations of such aggregated values remain very limited, with a mean of about $0.018$ in both cases, and a $75{th}$ percentile equal to~$0$, indicating that for the majority of instances models predicted the same value across runs, despite a small number of outliers. In practice, we found the ranking induced by the generated scores to be highly stable across independent runs.

\renewcommand{\arraystretch}{1.3}

\begin{table}[]
    \centering
    \footnotesize

    \caption{Distribution of propaganda techniques in tweets and news spans.}
    \label{tab:propaganda_tech_numbers}

    \setlength{\tabcolsep}{3pt}
    \scalebox{1}{
    \begin{tabular}{lrr}

    \toprule
    
    \multicolumn{1}{l}{\textbf{Technique}} & 
    \multicolumn{1}{c}{\textbf{N. tweets}} & 
    \multicolumn{1}{c}{\textbf{N. news spans}} 
    \\

    \midrule
     
    \textbf{No persuasion} 
    & 25,559 & 6,932 \\

    \textbf{Slogans} 
    & 1,584 & 71 \\
    
    \textbf{Name calling/labeling} 
    & 1,071 & 428 \\

    \textbf{Loaded language} 
    & 734 & 1,088 \\
    
    \textbf{Appeal to fear/prejudice}
    & 492 & 179\\

    \textbf{Doubt}
    & 64 &  318\\
    
    \textbf{Reductio ad Hitlerium} 
    & 50 & -\\

    \textbf{Bandwagon}
    & 25 & - \\
    
    \textbf{Flag waving} 
    & 13 & 136 \\

    \textbf{Black and white fallacy}
    & 1  & - \\
    
    \textbf{Appeal to authority} 
    & 1  & 82 \\

    \textbf{Thought terminating cliches} 
    & 1 & -  \\

    \textbf{Whataboutism} 
    & 1  & 4 \\

    \textbf{Repetition} 
    & -  & 272 \\

    \textbf{Exaggeration - Minimization} 
    & -  & 142 \\

    \textbf{False dilemma - No choice} 
    & -  & 75 \\

    \textbf{Causal oversimplification} 
    & -  & 71 \\

    \textbf{Conversation killer} 
    & -  & 53 \\

    \textbf{Red herring} 
    & -  & 28 \\

    \textbf{Appeal to popularity} 
    & -  & 22 \\

    \textbf{Guilt by association} 
    & -  & 19 \\

    \textbf{Appeal to hypocrisy} 
    & -  & 17 \\

    \textbf{Obfuscation - vagueness }  
    & -  & 16 \\

    \textbf{Straw man} 
    & -  & 12 \\

    \bottomrule
    
    \end{tabular}
    }

\end{table}

\section{Prompt Robustness}
\label{app:prompt_robustness}

To assess the robustness of our findings with respect to prompt wording, we replicated the experiments using an alternative formulation, which additionally inverted the direction of the Likert scale to introduce a stronger perturbation with respect to the original setup:

\begin{tcolorbox}[
    colback=gray!10,
    colframe=gray!50,
    arc=4pt,
    boxrule=0.5pt,
    left=6pt, right=6pt, top=6pt, bottom=6pt
]
\small
\textit{As an average social media user, to what extent would you support the following content online?}

\medskip
\textit{[CONTENT]}

\medskip
\textit{Please reply with a single number from 1 to 5:\\
1: Strongly support. \quad 2: Support. \quad 3: Neutral. \\ 4: Oppose. \quad 5: Strongly oppose.}
\end{tcolorbox}

Since the alternative prompt employed an inverted Likert scale, generated scores were reverse-coded before comparison with the original setup.

We then compared the endorsement scores produced by the two prompt variants. Across the tweets dataset, the mean absolute difference between the two formulations was 1.052, while 70.99\% of predictions differed by at most one point on the 5-point Likert scale. For the news dataset, the mean absolute difference was 0.809, with 80.04\% of predictions differing by at most one point.

\section{Geographical Origins of the Models}
\label{app:model_geographical_origin}
We examine whether and to what extent endorsement patterns align with or relate to the geographical origin of the models considered (cf. Table~\ref{tab:models}). Interestingly, while individual models exhibit different baseline endorsement tendencies, these do not map directly onto the model origin. For instance, in the case of the tweets dataset, Qwen shows one of the lowest baseline endorsements, whereas Yi does not follow the same trend, despite sharing the geographical origin. Similarly, for news, Qwen and Mistral show similar baseline endorsement despite being from different regions (cf. Appendix~\ref{app:endorsement_model}).
This finding suggests that endorsement behavior is more likely shaped by model-specific alignment and post-training choices~\cite{bernardelle2025mapping,han2025exploring} than by geographic background alone.

\section{Persuasion technique endorsement by Model}
\label{app:endorsement_model}
Model-specific baseline endorsement predictions vary across the technique-specific LMEs. Each distribution in Figures ~\ref{fig:boxplots_tweets} and ~\ref{fig:boxplots_news} represents the no-persuasion, neutral-persona baseline predictions obtained for an LLM across the four technique-specific models.
Baseline endorsement of different persuasion techniques under neutral persona prompting varies across models. As shown in Figure~\ref{fig:boxplots_tweets}, Mistral and Qwen exhibit notably lower baseline predictions than all other models for tweets. For news (Figure~\ref{fig:boxplots_news}), Mistral and Qwen similarly show lower baselines predictions, while Phi stands out in the opposite direction with a notably higher one.

\begin{figure}[]
    \centering
    \includegraphics[width=0.98\linewidth]{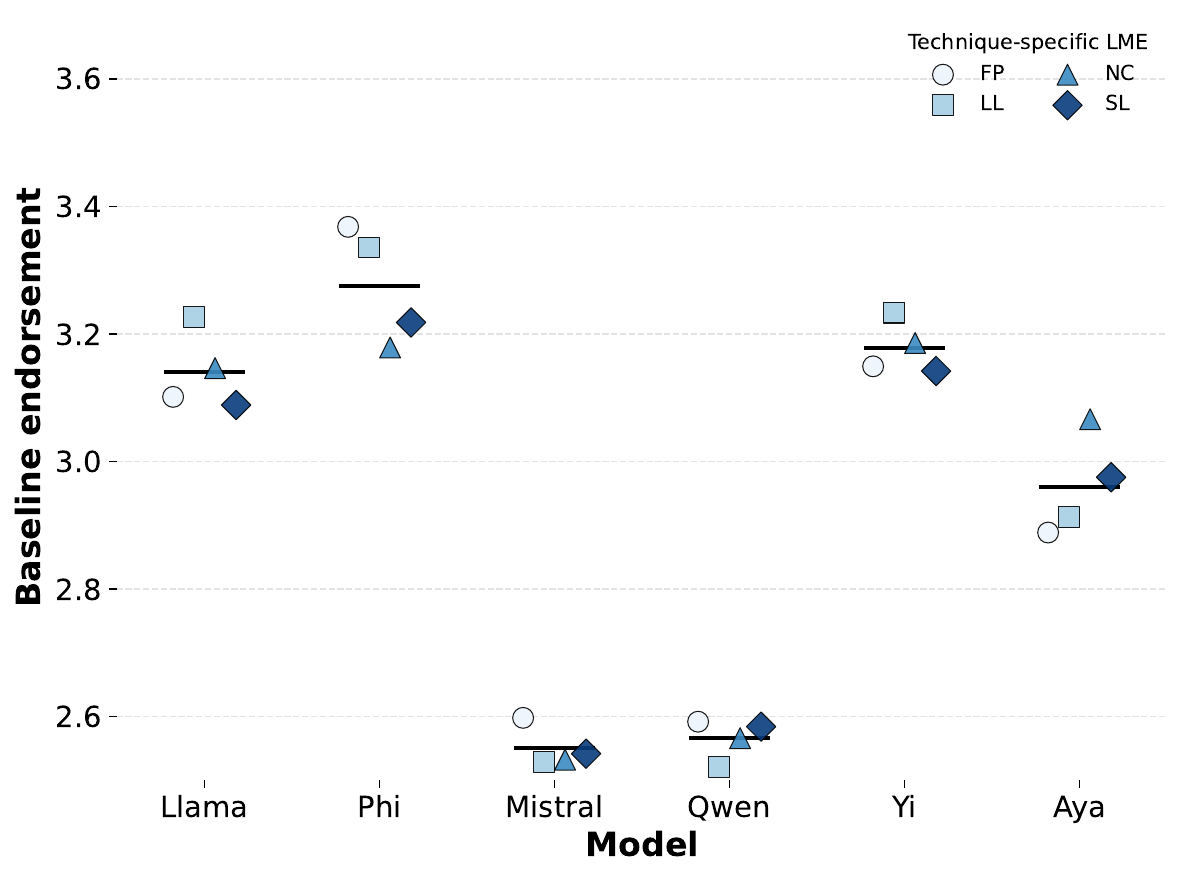}
    \caption{Distribution of model-specific baseline endorsement predictions across the four technique-specific LMEs for the tweet dataset}
    \label{fig:boxplots_tweets}
\end{figure}

\begin{figure}[]
    \centering
    \includegraphics[width=0.98\linewidth]{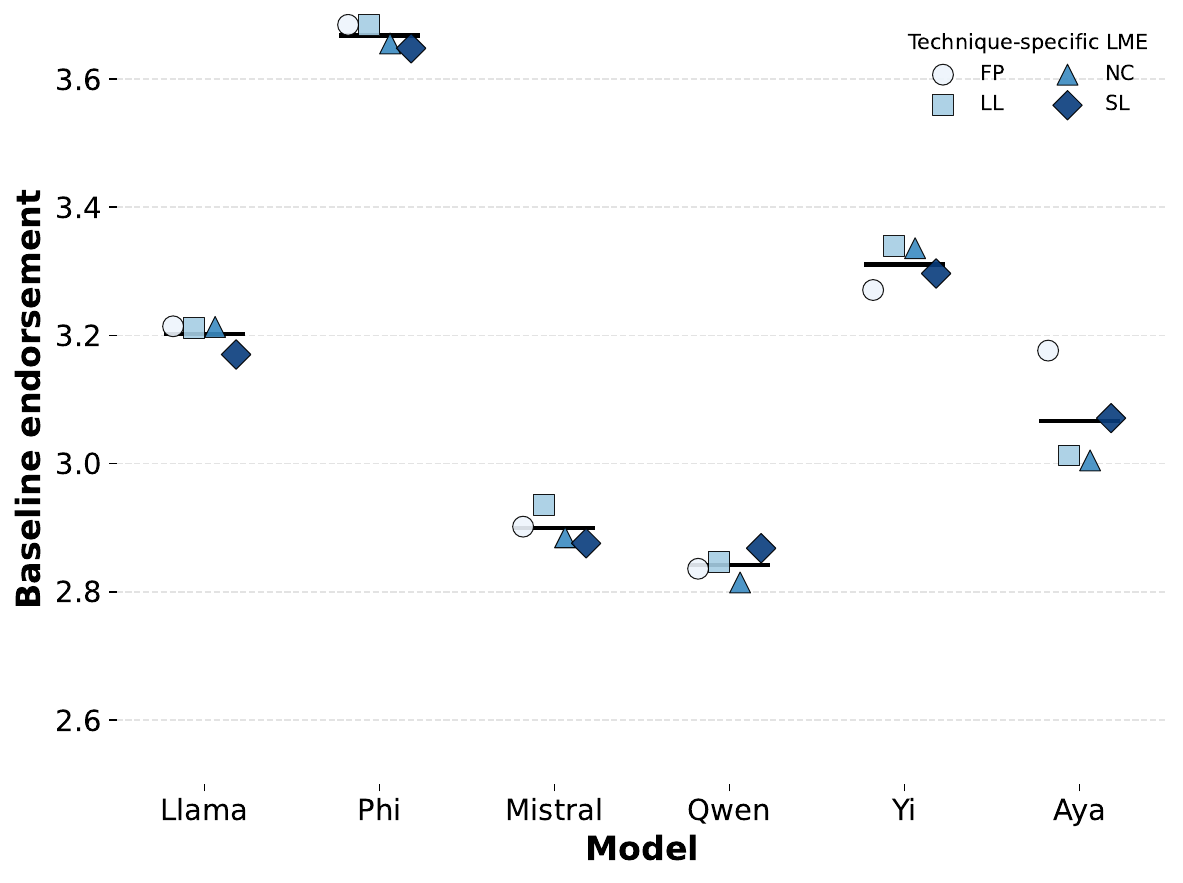}
    \caption{Distribution of model-specific baseline endorsement predictions across the four technique-specific LMEs for the news dataset.}
    \label{fig:boxplots_news}
\end{figure}

\section{Computing environment}\label{app:comp_env}
All models were deployed on an 8x NVIDIA A30 GPU server with 24 GB of RAM each, 764 GB of system RAM, a Double Intel Xeon Gold 6248R with a total of 96 cores, and Ubuntu Linux 20.04.6 LTS as OS.

\vspace{-3mm}
\section{Topic Naming}
\label{app:topic_naming}

We applied BERTopic~\cite{grootendorst2022bertopic} for topic modeling to each dataset, which groups documents into semantically coherent clusters based on their embeddings. Specifically, BERTopic was configured with a UMAP dimensionality reduction step using 30 neighbors, 5 components, cosine distance, and a minimum distance of 0.0. Clustering was performed with HDBSCAN, using a minimum cluster size of 75 for the tweet dataset and 100 for the news dataset. To make the resulting topics interpretable, we assigned a short descriptive label to each topic using \texttt{openai/gpt-4o-mini}. Specifically, for each topic, we provided the model with its ID, its top keywords, and a sample of representative documents, prompting it to return a concise label of at most two words, as shown next. \\ 

\vspace{-5mm}
\begin{tcolorbox}[
    colback=gray!10,
    colframe=gray!50,
    arc=4pt,
    boxrule=0.5pt,
    left=6pt, right=6pt, top=6pt, bottom=6pt
]
\small
\textit{I have a topic from a persuasion/LLM endorsement dataset.}

\medskip

\textit{Topic ID: [topicid]\\Topic keywords: [keywords]\\Example documents: [docs]}

\medskip

\textit{Based on the keywords and example documents, give a short label for the topic.}

\medskip

\textit{Rules:\\max 2 words\\no punctuation\\no explanation\\return only the label}

\medskip

\end{tcolorbox}

The resulting BERTopic clusters, their automatically generated labels, and their relative sizes are reported in Table~\ref{tab:topic_clusters}.

\begin{table*}[t!]
\centering
\small
\begin{tabular}{r l r|r l r}
\toprule
\multicolumn{3}{c|}{\textbf{Tweets}} & \multicolumn{3}{c}{\textbf{News}} \\
\textbf{ID} & \textbf{Topic} & \textbf{\% Docs} &
\textbf{ID} & \textbf{Topic} & \textbf{\% Docs} \\
\midrule
0  & Russia Ukraine         & 21.4 & 0  & Church Abuse          & 17.1 \\
1  & Ukrainian Nazis        & 12.2 & 1  & Mueller Investigation & 16.2 \\
2  & Ukrainian Propaganda   & 6.8  & 2  & Brexit Deal           & 16.2 \\
3  & Russian Propaganda     & 6.7  & 3  & Iran Regime           & 5.8 \\
4  & Media Propaganda       & 6.5  & 4  & Islam Israel          & 5.4 \\
5  & NATO Threat            & 5.8  & 5  & Court Allegations     & 5.1 \\
6  & Zelensky Criticism     & 5.8  & 6  & Public Discourse      & 5.1 \\
7  & Azov Nazis             & 5.1  & 7  & Las Vegas             & 3.8 \\
8  & Putin Hitler           & 4.4  & 8  & Permission Articles   & 3.4 \\
9  & Donbass Genocide       & 4.3  & 9  & Immigration Arrests   & 3.2 \\
10 & Nazi Propaganda        & 3.7  & 10 & Gun Control           & 2.7 \\
11 & Ukraine War            & 3.6  & 11 & Disease Outbreak      & 2.6 \\
12 & EU Sanctions           & 3.5  & 12 & Police Shootings      & 2.6 \\
13 & Biden Crisis           & 3.4  & 13 & White House           & 2.1 \\
14 & Trump DOJ              & 3.2  & 14 & Clinton Lies          & 2.0 \\
15 & China Taiwan           & 2.0  & 15 & London Protests       & 2.0 \\
16 & Nazi Ukraine           & 1.9  & 16 & Border Security       & 1.7 \\
   &                        &      & 17 & Facebook Censorship   & 1.5 \\
   &                        &      & 18 & Diplomatic Health     & 1.5 \\
\bottomrule
\end{tabular}
\caption{BERTopic clusters identified in the tweet and news datasets. The \textit{Topic} column reports automatically generated labels produced by \texttt{gpt-4o-mini}. Percentages indicate the proportion of topic-assigned documents belonging to each cluster.}
\label{tab:topic_clusters}
\end{table*}
\end{document}